\documentclass[twocolumn,prl]{revtex4}
\usepackage[T1]{fontenc}
\usepackage[latin9]{inputenc}
\usepackage{color}
\usepackage{amsmath}
\usepackage{amssymb}
\usepackage{graphicx}
\usepackage{esint}
\usepackage{amssymb}

\begin{document}

\title{Unsharp eigenvalues and quantum contextuality}
\author{F. De Zela}
\affiliation{Departamento de Ciencias, Secci\'{o}n F\'{i}sica,
Pontificia Universidad Cat\'{o}lica del Perú, Ap. 1761, Lima,
Peru.}

\begin{abstract}

The Kochen-Specker theorem, Bell inequalities, and several other
tests that were designed to rule out hidden-variable theories,
assume the existence of observables having infinitely sharp
eigenvalues. A paradigmatic example is spin-$1/2$. It is measured
with a Stern-Gerlach array whose outputs are divided into two
classes, spin-up and spin-down, in correspondence to the two spots
observed on a detection screen. The spot's finite size is
attributed to imperfections of the measuring device. This
assumption turns the experimental output into a dichotomic,
discrete one, thereby allowing the assignment of each spot to an infinitely sharp
eigenvalue. Alternatively, one can assume that the spot's finite
size stems from eigenvalues spanning a continuous range. Can we
disprove such an assumption? Can we rule out hidden-variable
theories that reproduce quantum predictions by assuming that, e.g., the
electron's magnetic moment is not exactly the same for all
electrons? We address these questions by focusing on the
Peres-Mermin version of the Bell-Kochen-Specker theorem. It is
shown that the assumption of unsharp eigenvalues precludes ruling
out non-contextual hidden-variable theories and hence quantum
contextuality does not arise. Analogous results hold for Bell-like
inequalities. This represents a new loophole that spoils several
fundamental tests of quantum mechanics and issues the challenge to
close it.

\end{abstract}


\maketitle

\section{Introductory remarks}

There is a fundamental prescription in quantum mechanics (QM) that
has been once qualified as a ``precept of the founders''
\cite{Mermin1}, namely the claim that it makes no sense to assign
values to unmeasured observables. Such a precept was turned into a
theorem by Bell \cite{Bell} and independently by Kochen and
Specker \cite{Kochen}, being since referred to as the
Bell-Kochen-Specker (BKS) theorem. It shows that it is impossible
to construct a non-contextual hidden-variable (HV) theory that
reproduces the predictions of QM. Non-contextuality means that the
results obtained by measuring an observable are independent of any
previous or simultaneous measurements on other, compatible
\cite{Szangolies} observables. The impossibility of constructing a
non-contextual HV theory is often expressed by saying that QM is
contextual. Now, such a feature of QM -- though being perhaps
somewhat peculiar -- does not seem to frontally collide with
common sense nor with possible approaches that might be undertaken
when pursuing scientific endeavors. Indeed, contextual models in
the social and in the natural sciences are perfectly acceptable
and imply no conflict with common sense. But the claim that we
cannot even assume that an observable has a value before it has
been measured is certainly at odds with common sense. Such a
feature sets QM apart from the rest of science. The idea that the
value of an observable comes into being just through its
measurement is something that conflicts with our most basic
notions of reality; a reality that keeps existing -- so we think
-- even if we do not interact with it. On the other hand, if we
assume that precise values can be assigned to unmeasured
observables, then we run into logical contradictions, as the BKS
theorem shows. Thus, it seems that we must pay a high price in
order to provide the quantum formalism with a self-consistent
ontology. This price is the abandonment of our most basic notions
of reality, something we are reluctant to do even as practitioners
of quantum physics. Classical ontology -- according to which
measurements of observables just reveal preexisting values -- must
be replaced by quantum ontology, if we want the quantum formalism
to be not merely a computational tool, but a consistent model of
the real world; a world in which we include ourselves, if
necessary, as perceiving subjects. Now, in spite of all these
needs we keep talking and thinking in terms of a classical
ontology. This hints at a latent conflict between the quantum
formalism and its interpretation in terms of our deeply rooted
notions of reality. The BKS theorem brought this conflict into
clearest light, and even more so the version of it due to Peres
and Mermin \cite{Mermin1,Peres}. The following conclusion seems
therefore to be unavoidable: we have to abandon the naive notion
of an external reality that exists independently of us. And yet,
this conclusion might be nonetheless avoidable. Indeed, let us
notice that in order to turn the aforementioned ``precept of the
founders'' into a logical consequence of the quantum formalism,
the BKS theorem had to invoke another ``precept of the founders''.
This precept states that some observables have infinitely sharp
eigenvalues. Any deviations from these sharply defined
(eigen)values should be attributable to measurement disturbances,
i.e., to imperfections of our measuring devices. Alas, the two
precepts seem to be in conflict with one another. For, first, we
are told to accept that the values of an observable are brought
into being by the very act of measurement. Thereafter, we are
asked to accept that the values we have recorded by measurement
need not always be the ``true'' ones. In most cases, so we are
told, measurements show values that only approximate the ``true''
ones. Why should we accept this statement without having any
compelling evidence of its truth? Paraphrasing Mermin
\cite{Mermin2}, we may perhaps say that it has been merely
reverence for the Patriarchs what diverted people from objecting a
precept that appears to be nothing but a misapplication of the
other, already accepted one. The word \emph{quantum} reminds us of
the strong appeal that sharp, integer values had during the
foundational period of QM. The impressive successes of the quantum
formalism surely helped to firmly establish the belief on sharp
eigenvalues as a mandatory prescription of the quantum creed. The
positivist commitment of the founders, which led them to deny the
very existence of what has not been measured, was curiously
betrayed by the founders themselves, who took for granted the
existence of discrete, infinitely sharp, ideal eigenvalues. If we
instead consistently rely on measurement outcomes alone, then we
have no reason to assume that observables must have infinitely
sharp eigenvalues.

\begin{figure}
\includegraphics[angle=0,scale=.5]{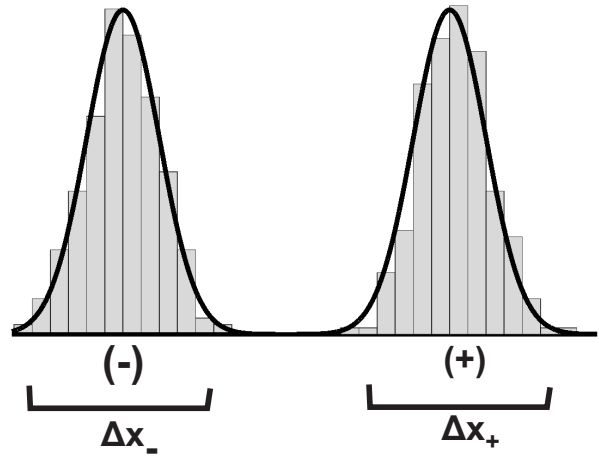}
\caption{Schematic representation of the detection part (detector screen) in a
Stern-Gerlach array. All spin-$1/2$ particles falling within the
spot labeled $\Delta x_{+}$ ($\Delta x_{-}$) are assigned to the
spin-up (spin-down) state. Under the assumption of infinitely sharp spin
values, spreads such as $\Delta x_{\pm}$ are attributed to
measurement uncertainties. Unsharp spin values could also
explain the observed results, similarly to cases in which one
measures classical observables -- e.g. weights -- of a population
containing some two varieties.}
\label{f1}
\end{figure}

Before we analyze the consequences of entertaining the rather
unusual assumption of unsharp eigenvalues, let us consider an
archetypical measurement, namely that of a spin-$1/2$.
Fig.~(\ref{f1}) shows schematically the detector part of a
Stern-Gerlach array (SGA). Particles in the spin-up state
$\left\vert \uparrow \right\rangle $ produce a click in the
(+)-detector, and correspondingly for the state $\left\vert
\downarrow \right\rangle $ and the
(-)-detector. Submitted to the action of the SGA, a spin-state $%
\left\vert \psi \right\rangle =a\left\vert \uparrow \right\rangle
+b\left\vert \downarrow \right\rangle $ is brought into a
spin-path entangled state: $\left\vert \psi \right\rangle
\rightarrow \left\vert \Psi \right\rangle =a\left\vert \uparrow
\right\rangle \left\vert +\right\rangle +b\left\vert \downarrow
\right\rangle \left\vert
-\right\rangle $, so that the probability that the (+)-detector fires is $%
p_{+}=\left\vert a\right\vert ^{2}$. Here, it is assumed that the
particle beam is well collimated, so that spin-up particles can
reach only the (+)-detector. Otherwise, the measurement is
unsharp. The SGA can be taken as representative of all
measurements. It is by reading some pointer that we fix the value
of whatever observable we want to measure. Any pointer
has a finite resolution, as it is illustrated in fig.~(\ref{f1}) by the lengths $%
\triangle x_{\pm}$. All particles being detected within $\triangle
x_{+}$ are assigned the infinitely sharp spin-value $+1$ (in units
of $\hbar /2$). Most particles fall around the middle of the
$\triangle x_{+}$ zone. The spatial spreading of the detected
particles is attributed to imperfections of the SGA, which
includes source and detector parts. Fig.~(\ref{f1}) shows two
fitted histograms. Let us assume for a moment that these
histograms correspond to macroscopic objects, apples of two
varieties, for example, grown in two different countries. Instead
of having recorded particles' positions $x$ we assume having
recorded the weights $w$ of apples in a sample that contains the
two varieties. Let $w_{\pm}$ be the two mean values of these
weights. If we weigh an apple of the sample and obtain, e.g.,
$w>w_{+}$, we do not interpret this outcome by saying that the
true value is not $w$ but $w_{+}$, and that any deviation from
$w_{+}$ must be attributed to an imperfect measurement. This is so
because
we can weigh the same apple many times, thereby obtaining values such as $%
w\pm\delta w$, with $\delta w\ll w$, that average out tightly at $w$. Imagine now that our measuring
procedure is such that in order to weigh an apple we must destroy
it. In such a case, nothing would prevent us from saying that the
spread in weights comes from imperfect measurements and that all
the apples in our sample are produced by nature with a weight that
is either $w_{+}$ or $w_{-}$. This is what happened in QM, which
originally dealt with microscopic objects that got destroyed when
submitted to measurement. Measurement's accuracy was assessed by
repeating the experiment on ``identically prepared'' replicas,
thereby taking for granted that, say, electrons are characterized
by sharply defined values. Quantum non-demolition measurements
were not available at that time. This kind of measurement is now
often applied \cite{Kirchmair}, although not with the aim of
testing the assumption of infinitely sharp eigenvalues. Anyhow, it
is clear that such an assumption is not the only possible one. It
might occur that, like apples, also particles possess spin values
that could slightly differ from their mean values $\pm \hbar /2$.
In the following, we will entertain the assumption that
observables have unsharp eigenvalues and study the consequences of
this assumption for some tests of quantum contextuality. As we
shall see, these consequences can be limited to HV models, leaving
QM untouched. The latter remains being what it always has been: an
ideal, extremely accurate model of physical reality. In this
model, observables are represented by operators whose sharply
defined eigenvalues coincide with the mean values of measured
observables.

\section{The Mermin-Peres version of the Bell-Kochen-Specker theorem}

Let us address now the Mermin-Peres version of the BKS theorem. It
will be convenient to use Mermin's first version of it
\cite{Mermin1}, that we reproduce here for completeness' sake and
future reference. This version applies to a four-dimensional
Hilbert space that corresponds to two qubits. We write, e.g.,
$X_{1}$ for $\sigma _{x}^{(1)}$, the Pauli $x$-matrix of the
first qubit. A HV theory ascribes the value $v(O)$ to the observable $%
O$. If a set of mutually commuting observables identically satisfy
a functional relationship $f(A,B,C,\ldots )=0$, then
this relationship must also be satisfied by the assigned values: $%
f(v(A),v(B),v(C),\ldots )=0$. Thus, it must hold
$v(AB)=v(A)v(B)$, whenever $\left[ A,B\right] =0$. By
considering operator identities such as
$(X_{1}Y_{2})(Y_{1}X_{2})(Z_{1}Z_{2})=I$, etc., one gets the
Mermin system of equations:
\begin{subequations}
\label{mermin}
\begin{eqnarray}
v(X_{1})v(X_{2})v(X_{1}X_{2}) &=&1,  \label{mermin1} \\
v(Y_{1})v(Y_{2})v(Y_{1}Y_{2}) &=&1,  \label{mermin2} \\
v(X_{1})v(Y_{2})v(X_{1}Y_{2}) &=&1,  \label{mermin3} \\
v(Y_{1})v(X_{2})v(Y_{1}X_{2}) &=&1,  \label{mermin4} \\
v(X_{1}Y_{2})v(Y_{1}X_{2})v(Z_{1}Z_{2}) &=&1,  \label{mermin5} \\
v(X_{1}X_{2})v(Y_{1}Y_{2})v(Z_{1}Z_{2}) &=&-1.  \label{mermin6}
\end{eqnarray}
\end{subequations}%
The above six equations cannot hold simultaneously. This claim is derived as follows \cite{Mermin1}: The assigned values $v(O)$ are
such that $v(O)\in \{-1,1\}$. This is so because in a HV-theory
$v(O)$ must be one of the possible measurement outcomes for $O$.
According to QM -- and, allegedly, to experimental evidence -- these outcomes
are $O$'s eigenvalues. Now, each value appears exactly twice on
the left of the above equations. Hence, the product of all values
on the left gives $1$. Since the product of the right sides is
$-1$, we get a contradiction. The assignment of values under the
above restrictions is thus impossible. Clearly, the restriction
$v(O)\in \{-1,1\}$ plays a key role. To substantiate it,
experimental evidence is often invoked. However, what
experimental evidence imposes is that $v(O)=\pm 1+\delta _{\pm }$,
for some $\delta _{\pm }$. Let us thus see the consequences of imposing
this last restriction instead of $v(O)=\pm 1$.

\section{A modified Mermin set of equations}

Of course, besides $v(O)=\pm 1+\delta _{\pm }$, we
must include some additional restrictions, e.g., that $X_{1}$ and $%
Y_{1}$ cannot be measured simultaneously. Moreover, all the above values $%
v(O)$ must have an operational meaning. Taking for example
eq.~(\ref{mermin1}), we
assume that it corresponds to an experimental array that is well suited for the measurement of $%
X_{1}$ and $X_{2}$. The value assigned to $X_{1}X_{2}$ is then given by $%
v(X_{1}X_{2})=v(X_{1})v(X_{2})$. Thus, we can consistently write $%
v(X_{1})v(X_{2})v(X_{1}X_{2})=v(X_{1})^{2}v(X_{2})^{2}.$ If we set $%
v(X_{1})=\pm 1+\delta _{x1}$ and $v(X_{2})=\pm 1+\delta _{x2}$, then $%
v(X_{1}X_{2})=\pm 1+\delta _{x1x2}$. The value of $\delta _{x1x2}$
follows from $\delta _{x1}$ and $\delta _{x2}$ in a way that the
HV-model should prescribe. We consider models for which the assignments
$v(X_{i})=\pm 1+\delta _{xi}$ reflect that spin values are
unsharp, i.e., spread around the mean values
$\pm 1$, very much like the weights of two apples' varieties. We thus set $%
\delta _{x1x2}=v_{1}\delta _{x2}+\delta _{x1}v_{2}$ and replace, e.g.,
eq.~(\ref{mermin1}) by $(v_{1}+\delta _{x1})^{2}(v_{2}+\delta
_{x2})^{2}=1+2v_{1}v_{2}\delta _{x1x2}$, with $v_{i}=\pm 1$,
$i=1,2$. In other words, we treat the $\delta _{xi}$ as deviations
from the corresponding mean values and apply for quantities like
$v(X_{1})v(X_{2})$ the rules of error propagation.
Proceeding in this way, instead of eqs.~(\ref{mermin}) we get the
following set of equations:
\begin{subequations}
\label{fdz}
\begin{eqnarray}
(v_{1}+\delta _{1})^{2}(v_{2}+\delta _{2})^{2} &=&1+2v_{1}v_{2}\delta _{12},
\label{fdz7} \\
(v_{3}+\delta _{3})^{2}(v_{4}+\delta _{4})^{2} &=&1+2v_{3}v_{4}\delta _{34},
\label{fdz8} \\
(v_{1}+\delta _{1})^{2}(v_{4}+\delta _{4})^{2} &=&1+2v_{1}v_{4}\delta _{14},
\label{fdz9} \\
(v_{3}+\delta _{3})^{2}(v_{2}+\delta _{2})^{2} &=&1+2v_{3}v_{2}\delta _{32},
\label{fdz10} \\
(w_{1}+\Delta _{1})^{2}(w_{2}+\Delta _{2})^{2} &=&1+2w_{1}w_{2}\Delta _{12},
\label{fdz11} \\
-(w_{1}+\Delta _{1})^{2}(w_{3}+\Delta _{3})^{2}
&=&-1+2w_{1}w_{3}\Delta _{13}.  \label{fdz12}
\end{eqnarray}
\end{subequations}%
Eqs.~(\ref{fdz7} -- \ref{fdz10}) involve the parameters $v_{j}$ and $\delta _{j}$, with $%
j=1,\ldots ,4$, whereas eqs.(\ref{fdz11}, \ref{fdz12}) involve the parameters $w_{j}$ and $%
\Delta _{j}$, with $j=1,2,3$. They are defined as follows: $%
v(X_{1})=v_{1}+\delta _{1}$, $v(X_{2})=v_{2}+\delta _{2}$, $%
v(Y_{1})=v_{3}+\delta _{3}$, $v(Y_{2})=v_{4}+\delta _{4}$; $%
v(Z_{1}Z_{2})=w_{1}+\Delta _{1}$, $v(X_{1}Y_{2})=w_{2}+\Delta _{2}$, $%
v(Y_{1}Y_{2})=w_{3}+\Delta _{3}$. Here, $v_{j}$ and $w_{j}$ take
on the values $\pm 1$, while $\delta _{j}$ and $\Delta _{j}$ are
free parameters that besides entering the above equations can be
required to satisfy additional
constraints, such as $\left\vert \delta _{j}\right\vert \leq \epsilon $ and $%
\left\vert \Delta _{j}\right\vert \leq \epsilon $, with $\epsilon
\ll 1$. Because $\delta _{jk}=v_{j}\delta _{k}+\delta _{k}v_{j}$
($j,k\in \{1,2,3,4\} $) and $\Delta _{jk}=w_{j}\Delta _{k}+\Delta
_{k}w_{j}$ ($j,k\in \{1,2,3\}$) we have more free parameters than
equations. In fact, for all possible choices of $v_{j}$ and
$w_{j}$ we can solve eqs.~(\ref{fdz7} -- \ref{fdz10}) by
expressing three of the $\delta _{j}$ in terms of the fourth, and solve
eqs.~(\ref{fdz11}, \ref{fdz12}) by expressing two of the $\Delta
_{j}$ in terms of the third. In other words, we can always obtain
values for the $\delta _{j}$ and
$\Delta _{j}$ so that they satisfy the above equations, alongside with $%
\left\vert \delta _{j}\right\vert \leq \epsilon $ and $\left\vert \Delta
_{j}\right\vert \leq \epsilon $. As an example, we set $v_{1}=-1$, $v_{2}=1$%
, $v_{3}=1$, $v_{4}=-1$, $\epsilon =10^{-3}$ and obtain, among
other choices, $\delta _{1}=0.887444\times 10^{-4}$, $\delta
_{2}=0.23779\times 10^{-4}$, $\delta _{3}=-0.63717\times 10^{-7}$
, $\delta _{4}=-0.23779\times 10^{-4}$, while setting $w_{1}=1$,
$w_{2}=-1$, $w_{3}=-1$, we obtain $\Delta
_{1}=-0.15470\times 10^{-3}$, $\Delta _{2}=-0.57722\times 10^{-3}$ and $%
\Delta _{3}=0.15469\times 10^{-3}$. We have thus exhibited a
consistent assignment of values for the set of observables
entering the Peres-Mermin version of the KBS theorem.

Let us stress that eqs.~(\ref{fdz}) follow from very general
assumptions. Indeed, while we have set $v(AB)=v(A)v(B)$ for some
commuting observables $A$, $B$, we have not assumed that such a
product rule holds for non-commuting observables. Had we done so,
then we would have run into contradictions. Indeed, from an
operator identity such as $\left[ \sigma _{x},\sigma _{y}\right]
=2i\sigma _{z}$, it would follow that $v(\sigma _{x})v(\sigma
_{y})-v(\sigma _{y})v(\sigma _{x})=0=2iv(\sigma _{z})$, which
cannot hold together with $v(\sigma _{z})=\pm 1$. For this reason,
we cannot consistently apply the product rule for all the
equations in the Mermin
system, eqs.~(\ref{mermin}). For example, we cannot set $v(X_{1}X_{2})=v(X_{1})v(X_{2})$ in eq.~(%
\ref{mermin1}) and simultaneously $v(Y_{1}Y_{2})=v(Y_{1})v(Y_{2})$ in eq.~(\ref{mermin2}%
). This is also not required when proving the Peres-Mermin
theorem. Note that while we refrain from applying the product
rule, this does not make the model contextual. We do not apply the
product rule because otherwise the model would be inconsistent.
For the very same reason we do not set for
$v(X_{1}X_{2})$ in eq.~(\ref{mermin6}) a value that derives from the values $%
v(X_{1})=v_{1}+\delta _{1}$ and $v(X_{2})=v_{2}+\delta _{2}$ entering eq.~(%
\ref{mermin1}) (cf. eq.~(\ref{fdz7})). Eqs.~(\ref{mermin5},\ref{mermin6}) are thus set apart from eqs.~(%
\ref{mermin1} -- \ref{mermin4}), in the sense that they are
related to quite different and
independent experimental arrays. Indeed, let us consider the observables $%
X_{1}Y_{2}$, $Y_{1}X_{2}$ and $Z_{1}Z_{2}$ entering eq.~%
(\ref{mermin6}). Because any one of them is the product of the
other two, we need to measure only two of them and then apply the
product rule. These two observables constitute a complete set of
commuting observables, i.e., by fixing their eigenvalues we fix
the corresponding common eigenvector. Written in terms of the
eigenvectors of Pauli-$Z$, i.e., $Z\left\vert \pm \right\rangle
=\pm
\left\vert \pm \right\rangle $, the eigenvectors of the above observables read $%
\left\vert \Phi ^{\pm }\right\rangle =(\pm i\left\vert
++\right\rangle +\left\vert --\right\rangle )/\sqrt{2}$,
$\left\vert \Psi ^{\pm }\right\rangle =(\pm i\left\vert
+-\right\rangle +\left\vert -+\right\rangle )/\sqrt{2}$. That is,
they constitute a Bell-like basis. If we want to measure, say,
$Z_{1}Z_{2}$ and $X_{1}Y_{2}$, we must set up an array that
performs projective measurements represented by the four
projectors $\left\vert \Phi ^{\pm }\right\rangle \left\langle \Phi
^{\pm }\right\vert $ and $\left\vert \Psi ^{\pm }\right\rangle
\left\langle \Psi ^{\pm }\right\vert $. If, for instance, the
detector $\left\vert \Phi ^{+}\right\rangle \left\langle \Phi
^{+}\right\vert $ fires, we make the assignments
$v(Z_{1}Z_{2})=+1$ and $v(X_{1}Y_{2})=-1$, while  we assign to the
third observable, $Y_{1}X_{2}$, a value that equals the product of the
measured ones: $v(Y_{1}X_{2})=v(Z_{1}Z_{2})v(X_{1}Y_{2})=-1$, and
so on. All this holds under the assumption of infinitely sharp
eigenvalues. Assuming instead unsharp
eigenvalues, we set $v(Z_{1}Z_{2})=1+\Delta _{1}$. Eqs.~(\ref{fdz}%
) refer to this case. In particular, eq.~(\ref{fdz12}) comes from
considering the identity $(Z_{1}Z_{2})(Y_{1}Y_{2})=-X_{1}X_{2}$
and from assuming that our measuring device projects onto the
common eigenvectors of $Z_{1}Z_{2}$ and $Y_{1}Y_{2}$.

He have thus derived eqs.~(\ref{fdz}) by assuming realism and
non-contextuality, besides unsharp eigenvalues. We introduced as
much free parameters ($\delta _{i=1,\ldots ,4}$, $\Delta
_{j=1,2,3}$) as these assumptions allow. One could wonder if
further constrains on these parameters could arise from the
uncertainty relations. We know that if we measure two
non-commuting observables such as $\sigma _{x}$ and $\sigma _{y}$
on identically prepared systems, the respective outcomes fulfill
uncertainty relations. The general form of these relations reads
$\left( \Delta A\right) ^{2}\left( \Delta B\right) ^{2}\geq \left[
\left\langle C\right\rangle ^{2}+\left\langle F\right\rangle
^{2}\right] /4$, with $\left[ A,B\right] =iC$,
$F=AB+BA-2\left\langle A\right\rangle \left\langle B\right\rangle
$ and $\left( \Delta A\right) ^{2}=\left\langle A^{2}\right\rangle
-\left\langle A\right\rangle ^{2}$. Setting $A=\sigma _{x}$,
$B=\sigma _{y}$ we get $\left\langle \sigma _{x}\right\rangle
^{2}+\left\langle \sigma _{y}\right\rangle ^{2}+\left\langle
\sigma _{z}\right\rangle ^{2}\leq 1$, a condition that is clearly
satisfied no matter which state, pure or mixed, is submitted to
measurement. In any case, this condition imposes no further
restrictions on the values of $\delta _{1}$ and $\delta _{3}$,
which correspond to $X_{1}$ and $Y_{1}$, respectively. Similar
considerations can be made for the other parameters.

In the special case when non-contextuality stems from locality --
i.e., if measurements are performed at spacelike separated
locations -- we can derive restricted forms of the
BKS-theorem that are expressed in terms of inequalities \cite{Mermin1}. A
well known one is the Clauser-Horne-Shimony-Holt (CHSH) inequality \cite%
{Clauser}. In contrast to the BKS-theorem, which holds for
arbitrary states, the CHSH-inequality holds for maximally
entangled states and involves four observables, $A_{1}$,
$A_{1}^{\prime }$, $B_{2}$ and $B_{2}^{\prime }$, whose
eigenvalues ($a_{1}$, $a_{1}^{\prime }$, $b_{2}$, $b_{2}^{\prime
}$) are $\pm 1$. Last restriction implies that
$(a_{1}+a_{1}^{\prime })b_{2}+(a_{1}-a_{1}^{\prime })b_{2}^{\prime
}=\pm 2$. From this, one readily derives the CHSH-inequality
$\left\vert \left\langle A_{1}B_{2}\right\rangle +\left\langle
A_{1}^{\prime }B_{2}\right\rangle +\left\langle A_{1}B_{2}^{\prime
}\right\rangle -\left\langle A_{1}^{\prime }B_{2}^{\prime
}\right\rangle \right\vert \leq 2$, which QM violates for
appropriate choices of the involved states and observables. The
assumption of infinitely sharp eigenvalues plays an essential role here as
well. By dropping it we should be able to explain any experimental
outcomes, as we have enough free parameters at our disposal.
Similar considerations should apply to other versions of Bell-like
inequalities and to different variants of the BKS theorem
\cite{Yu,Zu,Canhas}.

\section{Conclusions}

As we have seen, the assumption of unsharp eigenvalues has
far-reaching consequences for some fundamental tests of QM. A
related but quite different subject is that of unsharp
measurements. As an example of the latter we may refer to
disturbances that could restrict the compatibility of
observables being submitted to sequential measurements \cite%
{Szangolies,Kirchmair}. One can take these disturbances into
account and still produce results that HV models cannot explain
\cite{Kirchmair}. Our approach differs also from earlier ones that
addressed finite precision measurements \cite{Meyer,Kent}. It has
been shown that finite precision does not nullify the BKS theorem,
but rather hints at a different type of contextuality, called
``existential contextuality'' \cite{Appleby}. The consequences of
assuming unsharp eigenvalues seem to have been neglected. While
finite precision measurements might spoil our ability to meet the
benchmark set by fundamental tests of QM, the assumption of
unsharp eigenvalues spoils the benchmark itself. Unsharp
eigenvalues surely fit among the assumptions of HV theories and,
moreover, they are not alien to QM. Indeed, let us recall some
representative cases: Atomic energy spectra have discrete as well
as continuous -- i.e., unsharp -- parts, whereas in more complex
systems such as semiconductors one often deals with energy bands.
Faced with the natural linewidth of spontaneously emitted light,
one realizes that atomic energy states in the discrete part of the
spectrum cannot be infinitely sharp. The spread $\Delta E$ of a
level can be traced back to the coupling between atomic electrons
and electromagnetic fields that have continuous energy spectra.
The coupling can then modify an otherwise discrete part of the
spectrum. Moreover, this coupling involves electron's charge as
much as its magnetic moment $\mu $. Elementary particles are not
characterized by a fixed charge's value, as it was originally
assumed. Since long, ``running coupling constants'' are routinely
employed in high-energy physics. Hence, it is not physically
unreasonable to assume a spread $\Delta \mu $, which in turn
implies a spread of spin's eigenvalues. But independently of any
plausibility arguments, the fact is that there is a spread of
recorded values, which may be attributed to the quantity being
measured rather than to imperfections of the measuring procedure.
A consistent realist theory may be built upon such an assumption.
The BKS ban does not apply under such circumstances, and values
can be assigned to observables without running into
contradictions. We could however hope to rule out HV theories by
addressing observables whose eigenvalues span a continuous range.
In fact, CHSH-like inequalities have been derived for such a case,
as for instance in \cite{Borges}, where eigenvalues are given by
$\cos\theta$, with $\theta \in [0,\pi]$. Now, inequalities follow
from the fact that such eigenvalues are bounded: $|\cos\theta|\leq
1$. Thus, the role that was previously played by infinitely sharp
eigenvalues, is now played by an infinitely sharp boundary.
Clearly, a HV model could here again be constructed upon the
assumption that the boundary is unsharp, an assumption that would
be surely in agreement with experimental facts.

Finally, let us notice that foundational issues such as those
discussed in this work might be relevant for quantum information
theory as well \cite{Nagata,Aharon}. Any quantum device works with
inherent uncertainties of the kind illustrated by the SGA we have
considered here. Hence, classical analogs that mimic unsharp
quantum eigenvalues could shed light on several issues of current interest \cite{Buhrman}. Of particular relevance in this respect is the recent identification of quantum contextuality as a critical resource for quantum speed-up of fault-tolerant quantum computation \cite{Howard}.

%\acknowledgments
%This work was partially supported by DGI-PUCP, under Grant 2014-0064.


\begin{thebibliography}{99}

\bibitem{Mermin1} N. D. Mermin, Phys. Rev. Lett. 65 (1990) 3373.

\bibitem{Bell} J. S. Bell, Rev. Mod. Phys. 38 (1966) 447.

\bibitem{Kochen} S. Kochen and E. P. Specker, J. Math. Mech. 17 (1967) 59.

\bibitem{Szangolies} J. Szangolies, M. Kleinmann, and O. Gühne, Phys. Rev. A 87 (2013) 050101(R).

\bibitem{Peres} A. Peres, J. Phys. A 24 (1991) L175.

\bibitem{Mermin2} N. D. Mermin, Rev. Mod. Phys. 65 (1993) 803.

\bibitem{Kirchmair} G. Kirchmair \textit{et al.}, Nature (London) 460 (2009) 494.

\bibitem{Clauser} J. F. Clauser, M. A. Horne, A. Shimony, and R. A. Holt,
Phys. Rev. Lett. 23 (1969) 880.

\bibitem{Yu} S. Yu and C. H. Oh, Phys. Rev. Lett. 108 (2012)
030402.

\bibitem{Zu} C. Zu \textit{et al.}, Phys. Rev. Lett. 109 (2012) 150401.

\bibitem{Canhas} G. Cañas \textit{et al.}, Phys. Rev. A 90 (2014) 012119.

\bibitem{Meyer} D. A. Meyer, Phys. Rev. Lett. 83 (1999) 3751.

\bibitem{Kent} A. Kent, Phys. Rev. Lett. 83 (1999) 3755.

\bibitem{Appleby} D. M. Appleby, Phys. Rev. A 65 (2002) 022105.

\bibitem{Borges} C. V. S. Borges, P. Milman, and A. Keller, Phys.
Rev. A 86 (2012) 052107.

\bibitem{Nagata} K. Nagata, Phys. Rev. A 52 (2005) 012325.

\bibitem{Aharon} N. Aharon and L. Vaidman, Phys. Rev. A 77 (2008) 052310.

\bibitem{Buhrman} H. Buhrman \textit{et al.}, Rev. Mod. Phys. 82 (2010) 665.

\bibitem{Howard} M. Howard \textit{et al.}, Nature (London) 510 (2014) 351.

%\bibitem{Lacour} B. R. La Cour, Phys. Rev. A \textbf{79}, 012102 (2009).

\end{thebibliography}
\end{document}